%% file: paper.tex
\documentclass[letterpaper,compsoc,twoside]{IEEEtran}
\usepackage{fixltx2e} 
\usepackage{cmap} 
\usepackage{ifthen}
\usepackage[T1]{fontenc}
\usepackage[utf8]{inputenc}

\usepackage[font={small,it},labelfont=bf]{caption}
\usepackage{float}

\pdfoutput=1
\usepackage{scipy}
\makeatletter
\def\PY@reset{\let\PY@it=\relax \let\PY@bf=\relax%
    \let\PY@ul=\relax \let\PY@tc=\relax%
    \let\PY@bc=\relax \let\PY@ff=\relax}
\def\PY@tok#1{\csname PY@tok@#1\endcsname}
\def\PY@toks#1+{\ifx\relax#1\empty\else%
    \PY@tok{#1}\expandafter\PY@toks\fi}
\def\PY@do#1{\PY@bc{\PY@tc{\PY@ul{%
    \PY@it{\PY@bf{\PY@ff{#1}}}}}}}
\def\PY#1#2{\PY@reset\PY@toks#1+\relax+\PY@do{#2}}

\expandafter\def\csname PY@tok@gd\endcsname{\def\PY@tc##1{\textcolor[rgb]{0.63,0.00,0.00}{##1}}}
\expandafter\def\csname PY@tok@gu\endcsname{\let\PY@bf=\textbf\def\PY@tc##1{\textcolor[rgb]{0.50,0.00,0.50}{##1}}}
\expandafter\def\csname PY@tok@gt\endcsname{\def\PY@tc##1{\textcolor[rgb]{0.00,0.27,0.87}{##1}}}
\expandafter\def\csname PY@tok@gs\endcsname{\let\PY@bf=\textbf}
\expandafter\def\csname PY@tok@gr\endcsname{\def\PY@tc##1{\textcolor[rgb]{1.00,0.00,0.00}{##1}}}
\expandafter\def\csname PY@tok@cm\endcsname{\let\PY@it=\textit\def\PY@tc##1{\textcolor[rgb]{0.25,0.50,0.56}{##1}}}
\expandafter\def\csname PY@tok@vg\endcsname{\def\PY@tc##1{\textcolor[rgb]{0.73,0.38,0.84}{##1}}}
\expandafter\def\csname PY@tok@m\endcsname{\def\PY@tc##1{\textcolor[rgb]{0.13,0.50,0.31}{##1}}}
\expandafter\def\csname PY@tok@mh\endcsname{\def\PY@tc##1{\textcolor[rgb]{0.13,0.50,0.31}{##1}}}
\expandafter\def\csname PY@tok@cs\endcsname{\def\PY@tc##1{\textcolor[rgb]{0.25,0.50,0.56}{##1}}\def\PY@bc##1{\setlength{\fboxsep}{0pt}\colorbox[rgb]{1.00,0.94,0.94}{\strut ##1}}}
\expandafter\def\csname PY@tok@ge\endcsname{\let\PY@it=\textit}
\expandafter\def\csname PY@tok@vc\endcsname{\def\PY@tc##1{\textcolor[rgb]{0.73,0.38,0.84}{##1}}}
\expandafter\def\csname PY@tok@il\endcsname{\def\PY@tc##1{\textcolor[rgb]{0.13,0.50,0.31}{##1}}}
\expandafter\def\csname PY@tok@go\endcsname{\def\PY@tc##1{\textcolor[rgb]{0.20,0.20,0.20}{##1}}}
\expandafter\def\csname PY@tok@cp\endcsname{\def\PY@tc##1{\textcolor[rgb]{0.00,0.44,0.13}{##1}}}
\expandafter\def\csname PY@tok@gi\endcsname{\def\PY@tc##1{\textcolor[rgb]{0.00,0.63,0.00}{##1}}}
\expandafter\def\csname PY@tok@gh\endcsname{\let\PY@bf=\textbf\def\PY@tc##1{\textcolor[rgb]{0.00,0.00,0.50}{##1}}}
\expandafter\def\csname PY@tok@ni\endcsname{\let\PY@bf=\textbf\def\PY@tc##1{\textcolor[rgb]{0.84,0.33,0.22}{##1}}}
\expandafter\def\csname PY@tok@nl\endcsname{\let\PY@bf=\textbf\def\PY@tc##1{\textcolor[rgb]{0.00,0.13,0.44}{##1}}}
\expandafter\def\csname PY@tok@nn\endcsname{\let\PY@bf=\textbf\def\PY@tc##1{\textcolor[rgb]{0.05,0.52,0.71}{##1}}}
\expandafter\def\csname PY@tok@no\endcsname{\def\PY@tc##1{\textcolor[rgb]{0.38,0.68,0.84}{##1}}}
\expandafter\def\csname PY@tok@na\endcsname{\def\PY@tc##1{\textcolor[rgb]{0.25,0.44,0.63}{##1}}}
\expandafter\def\csname PY@tok@nb\endcsname{\def\PY@tc##1{\textcolor[rgb]{0.00,0.44,0.13}{##1}}}
\expandafter\def\csname PY@tok@nc\endcsname{\let\PY@bf=\textbf\def\PY@tc##1{\textcolor[rgb]{0.05,0.52,0.71}{##1}}}
\expandafter\def\csname PY@tok@nd\endcsname{\let\PY@bf=\textbf\def\PY@tc##1{\textcolor[rgb]{0.33,0.33,0.33}{##1}}}
\expandafter\def\csname PY@tok@ne\endcsname{\def\PY@tc##1{\textcolor[rgb]{0.00,0.44,0.13}{##1}}}
\expandafter\def\csname PY@tok@nf\endcsname{\def\PY@tc##1{\textcolor[rgb]{0.02,0.16,0.49}{##1}}}
\expandafter\def\csname PY@tok@si\endcsname{\let\PY@it=\textit\def\PY@tc##1{\textcolor[rgb]{0.44,0.63,0.82}{##1}}}
\expandafter\def\csname PY@tok@s2\endcsname{\def\PY@tc##1{\textcolor[rgb]{0.25,0.44,0.63}{##1}}}
\expandafter\def\csname PY@tok@vi\endcsname{\def\PY@tc##1{\textcolor[rgb]{0.73,0.38,0.84}{##1}}}
\expandafter\def\csname PY@tok@nt\endcsname{\let\PY@bf=\textbf\def\PY@tc##1{\textcolor[rgb]{0.02,0.16,0.45}{##1}}}
\expandafter\def\csname PY@tok@nv\endcsname{\def\PY@tc##1{\textcolor[rgb]{0.73,0.38,0.84}{##1}}}
\expandafter\def\csname PY@tok@s1\endcsname{\def\PY@tc##1{\textcolor[rgb]{0.25,0.44,0.63}{##1}}}
\expandafter\def\csname PY@tok@gp\endcsname{\let\PY@bf=\textbf\def\PY@tc##1{\textcolor[rgb]{0.78,0.36,0.04}{##1}}}
\expandafter\def\csname PY@tok@sh\endcsname{\def\PY@tc##1{\textcolor[rgb]{0.25,0.44,0.63}{##1}}}
\expandafter\def\csname PY@tok@ow\endcsname{\let\PY@bf=\textbf\def\PY@tc##1{\textcolor[rgb]{0.00,0.44,0.13}{##1}}}
\expandafter\def\csname PY@tok@sx\endcsname{\def\PY@tc##1{\textcolor[rgb]{0.78,0.36,0.04}{##1}}}
\expandafter\def\csname PY@tok@bp\endcsname{\def\PY@tc##1{\textcolor[rgb]{0.00,0.44,0.13}{##1}}}
\expandafter\def\csname PY@tok@c1\endcsname{\let\PY@it=\textit\def\PY@tc##1{\textcolor[rgb]{0.25,0.50,0.56}{##1}}}
\expandafter\def\csname PY@tok@kc\endcsname{\let\PY@bf=\textbf\def\PY@tc##1{\textcolor[rgb]{0.00,0.44,0.13}{##1}}}
\expandafter\def\csname PY@tok@c\endcsname{\let\PY@it=\textit\def\PY@tc##1{\textcolor[rgb]{0.25,0.50,0.56}{##1}}}
\expandafter\def\csname PY@tok@mf\endcsname{\def\PY@tc##1{\textcolor[rgb]{0.13,0.50,0.31}{##1}}}
\expandafter\def\csname PY@tok@err\endcsname{\def\PY@bc##1{\setlength{\fboxsep}{0pt}\fcolorbox[rgb]{1.00,0.00,0.00}{1,1,1}{\strut ##1}}}
\expandafter\def\csname PY@tok@kd\endcsname{\let\PY@bf=\textbf\def\PY@tc##1{\textcolor[rgb]{0.00,0.44,0.13}{##1}}}
\expandafter\def\csname PY@tok@ss\endcsname{\def\PY@tc##1{\textcolor[rgb]{0.32,0.47,0.09}{##1}}}
\expandafter\def\csname PY@tok@sr\endcsname{\def\PY@tc##1{\textcolor[rgb]{0.14,0.33,0.53}{##1}}}
\expandafter\def\csname PY@tok@mo\endcsname{\def\PY@tc##1{\textcolor[rgb]{0.13,0.50,0.31}{##1}}}
\expandafter\def\csname PY@tok@mi\endcsname{\def\PY@tc##1{\textcolor[rgb]{0.13,0.50,0.31}{##1}}}
\expandafter\def\csname PY@tok@kn\endcsname{\let\PY@bf=\textbf\def\PY@tc##1{\textcolor[rgb]{0.00,0.44,0.13}{##1}}}
\expandafter\def\csname PY@tok@o\endcsname{\def\PY@tc##1{\textcolor[rgb]{0.40,0.40,0.40}{##1}}}
\expandafter\def\csname PY@tok@kr\endcsname{\let\PY@bf=\textbf\def\PY@tc##1{\textcolor[rgb]{0.00,0.44,0.13}{##1}}}
\expandafter\def\csname PY@tok@s\endcsname{\def\PY@tc##1{\textcolor[rgb]{0.25,0.44,0.63}{##1}}}
\expandafter\def\csname PY@tok@kp\endcsname{\def\PY@tc##1{\textcolor[rgb]{0.00,0.44,0.13}{##1}}}
\expandafter\def\csname PY@tok@w\endcsname{\def\PY@tc##1{\textcolor[rgb]{0.73,0.73,0.73}{##1}}}
\expandafter\def\csname PY@tok@kt\endcsname{\def\PY@tc##1{\textcolor[rgb]{0.56,0.13,0.00}{##1}}}
\expandafter\def\csname PY@tok@sc\endcsname{\def\PY@tc##1{\textcolor[rgb]{0.25,0.44,0.63}{##1}}}
\expandafter\def\csname PY@tok@sb\endcsname{\def\PY@tc##1{\textcolor[rgb]{0.25,0.44,0.63}{##1}}}
\expandafter\def\csname PY@tok@k\endcsname{\let\PY@bf=\textbf\def\PY@tc##1{\textcolor[rgb]{0.00,0.44,0.13}{##1}}}
\expandafter\def\csname PY@tok@se\endcsname{\let\PY@bf=\textbf\def\PY@tc##1{\textcolor[rgb]{0.25,0.44,0.63}{##1}}}
\expandafter\def\csname PY@tok@sd\endcsname{\let\PY@it=\textit\def\PY@tc##1{\textcolor[rgb]{0.25,0.44,0.63}{##1}}}

\def\PYZus{\char`\_}
\def\PYZob{\char`\{}
\def\PYZcb{\char`\}}

\def\PYZsh{\char`\#}

\def\PYZhy{\char`\-}
\def\PYZsq{\char`\'}
\def\PYZdq{\char`\"}

\makeatother

\providecommand*{\DUrole}[2]{%
  \ifcsname DUrole#1\endcsname%
    \csname DUrole#1\endcsname{#2}%
  \else
    \ifcsname docutilsrole#1\endcsname%
      \csname docutilsrole#1\endcsname{#2}%
    \else%
      #2%
    \fi%
  \fi%
}

\ifthenelse{\isundefined{\hypersetup}}{
  \usepackage[colorlinks=true,linkcolor=blue,urlcolor=blue]{hyperref}
  \urlstyle{same} 
}{}

\begin{document}
\newcounter{footnotecounter}\title{JyNI – Using native CPython-Extensions in Jython}\author{Stefan Richthofer$^{\setcounter{footnotecounter}{1}\fnsymbol{footnotecounter}\setcounter{footnotecounter}{2}\fnsymbol{footnotecounter}}$%
          \setcounter{footnotecounter}{1}\thanks{\fnsymbol{footnotecounter} %
          Corresponding author: \protect\href{mailto:stefan.richthofer@gmx.de}{stefan.richthofer@gmx.de}}\setcounter{footnotecounter}{2}\thanks{\fnsymbol{footnotecounter} Institute for Neural Computation, Ruhr-Universität Bochum}\thanks{%

          \noindent%
          Copyright\,\copyright\,2014 Stefan Richthofer. This is an open-access article distributed under the terms of the Creative Commons Attribution License, which permits unrestricted use, distribution, and reproduction in any medium, provided the original author and source are credited. http://creativecommons.org/licenses/by/3.0/%
        }}\maketitle
          \renewcommand{\leftmark}{PROC. OF THE 6th EUR. CONF. ON PYTHON IN SCIENCE (EUROSCIPY 2013)}
          \renewcommand{\rightmark}{JYNI – USING NATIVE CPYTHON-EXTENSIONS IN JYTHON}

\InputIfFileExists{page_numbers.tex}{}{}
\newcommand*{\docutilsroleref}{\ref}
\newcommand*{\docutilsrolelabel}{\label}
\AtEndDocument{\cleardoublepage}
\begin{abstract}Jython is a Java based Python implementation and the most
seamless way to integrate Python and Java. However, it does
not support native extensions written for CPython like NumPy
or SciPy. Since most scientific Python code fundamentally
depends on exactly such native extensions directly or indirectly,
it usually cannot be run with Jython. JyNI (Jython Native Interface)
aims to close this gap. It is a layer that enables Jython users to
load native CPython extensions and access them from Jython the
same way as they would do in CPython. In order to leverage the JyNI
functionality, you just have to put it on the Java classpath when
Jython is launched. It neither requires you to recompile the
extension code, nor to build a customized Jython fork.
That means, it is binary compatible with existing extension builds.

At the time of writing, JyNI does not fully implement
the Python C-API and it is only capable of loading simple examples
that only involve most basic built-in types. The concept is rather complete
though and our goal is to provide the C-API needed to load NumPy as soon
as possible. After that we will focus on SciPy and others.

We expect that our work will also enable Java developers to use
CPython extensions like NumPy in their Java code.\end{abstract}

\section{Introduction%
  \label{introduction}%
}

\cite{JyNI} is a compatibility layer with the goal to enable
\cite{JYTHON} to use native CPython extensions like NumPy
or SciPy. This way we aim to enable scientific Python
code to run on Jython.\begin{figure}[]\noindent\makebox[\columnwidth][c]{\includegraphics[scale=0.30]{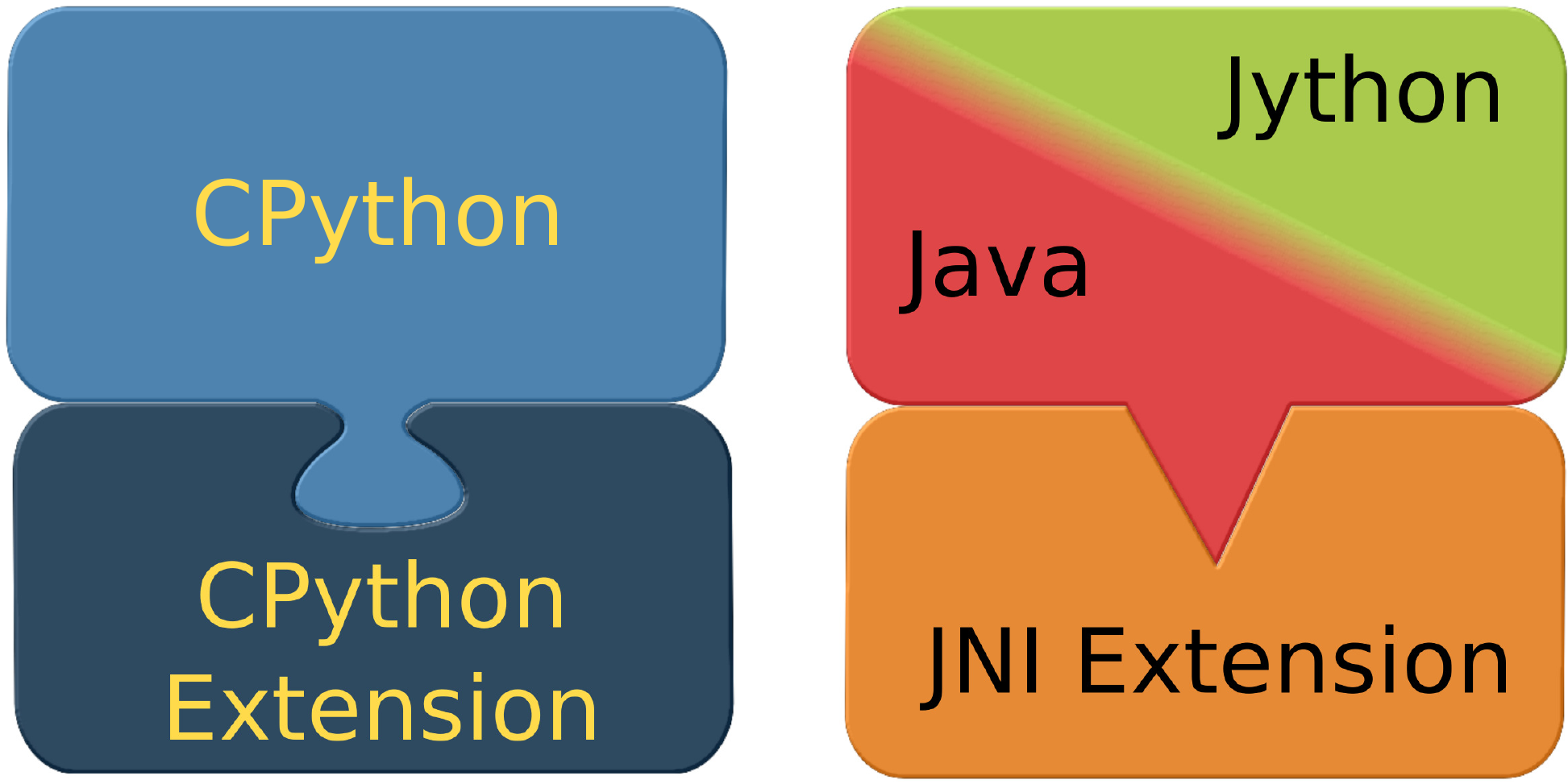}}
\caption{Extensions}
\end{figure}

Since Java is rather present in industry, while Python
is more present in science, JyNI is an important step
to lower the cost of using scientific code in industrial
environments.\begin{figure}[]\noindent\makebox[\columnwidth][c]{\includegraphics[scale=0.30]{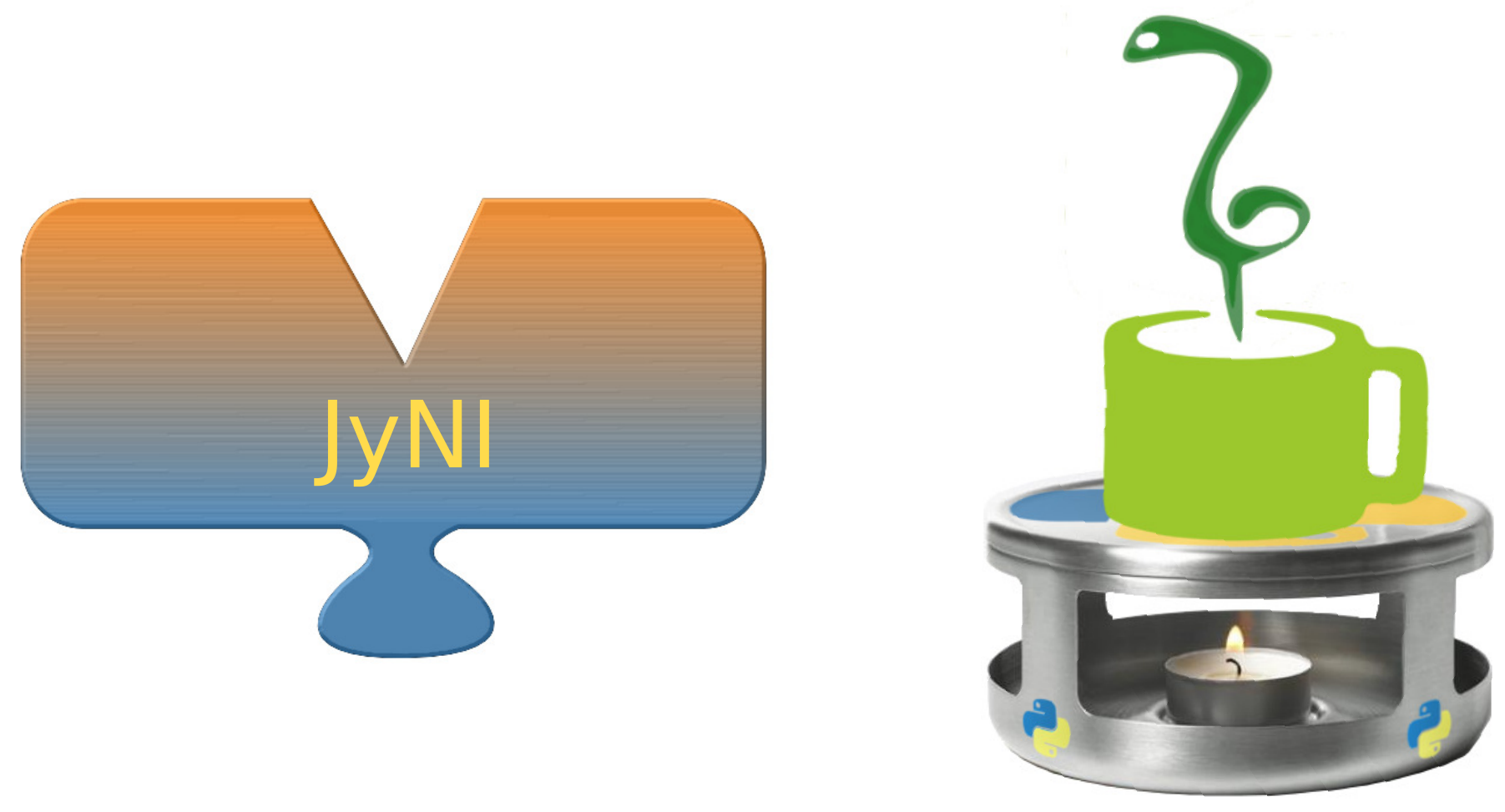}}
\caption{JyNI}
\end{figure}

Because of the complexity of the Python C-API, the task of developing JyNI revealed to be a true challenge.
Especially the frequent occurrence of preprocessor macros in the public C-API allows for no naive approaches like directly delegating every C-API call to Jython.
It turned out, that most built-in types need individual fixes, considerations and adjustments to allow seamless integration with Jython.

There have been similar approaches for other Python implementations, namely \cite{IRONCLAD} for IronPython and \cite{CPYEXT} for PyPy.
As far as we know, these suffer from equal difficulties as JyNI and also did not yet reach a satisfying compatibility level for
current Python versions.

Another interesting work is NumPy4J \cite{NP4J}, which provides Java interfaces for NumPy by embedding the CPython interpreter.
It features automatic conversion to Java suitable types and thus allows easy integration with other Java frameworks.
A more general approach is Jepp \cite{JEPP}, which also works via embedding the CPython interpreter.
It also includes conversion methods between basic Python and Java types, but is not specifically NumPy-aware.

However, none of these approaches aims for integration with Jython. In contrast to that, JyNI is entirely based on Jython.
Though large parts are derived from CPython, the main Python runtime is provided by Jython and JyNI delegates most C-API calls
to Jython directly or indirectly (i.e. some objects are mirrored natively, so calls to these can be processed entirely on native side, syncing the results with Jython afterwards; see implementation section for details).

\section{Usage%
  \label{usage}%
}

Thanks to Jython's hooking capabilities, it is sufficient to place \texttt{JyNI.jar} on the classpath (and some native libraries on the library path) when Jython is launched.
Then Jython should “magically” be able to load native extensions, as far as the needed Python C-API is already implemented by JyNI.
No recompilation, no forking – it just works with original Jython and original extensions (up to version compatibility; see the versioning notes at the end of this section).

Note that  the naive way does not actually set the classpath for Jython:%
\begin{quote}\begin{verbatim}
java -cp "[...]:JyNI.jar:[JyNI binaries folder]"
   -jar jython.jar
\end{verbatim}

\end{quote}
The correct way is:%
\begin{quote}\begin{verbatim}
java -cp "[...]:JyNI.jar:[JyNI binaries folder]:
   jython.jar" org.python.util.jython
\end{verbatim}

\end{quote}
Alternatively, one can use Jython's start script:%
\begin{quote}\begin{verbatim}
jython -J-cp "[...]:JyNI.jar:[JyNI binaries folder]"
\end{verbatim}

\end{quote}
Usually the JVM does not look for native library files on the classpath.
To ease the configuration, we built into JyNI's initializer code that it also searches for
native libraries on the classpath. Alternatively you can place \texttt{libJyNI.so} and
\texttt{libJyNI-loader.so} anywhere the JVM finds them, i.e. on the java library path (\texttt{java.library.path}) or the system's library path (\texttt{LD\_LIBRARY\_PATH}).

To get an impression of JyNI, proceed as described in the following subsection.

\subsection{Instructions to run \texttt{JyNIDemo.py}%
  \label{instructions-to-run-jynidemo-py}%
}
\begin{itemize}

\item 

Go to \cite{JyNI}, select the newest release in the download section and get the sources and binaries appropriate for your system (32 or 64 bit).
\item 

Extract \texttt{JyNI-Demo/src/JyNIDemo.py} from the sources.
\item 

To launch it with CPython, extract \texttt{DemoExtension.so} from the bin archive.
\item 

\texttt{JyNIDemo.py} adds the extension folder via \texttt{sys.path.append({[}path{]})}.
You can modify that line so it finds your extracted \texttt{DemoExtension.so} or delete the line and put
\texttt{DemoExtension.so} on the pythonpath.
\item 

If you launch \texttt{JyNIDemo.py} with Jython, it won't work.
Put \texttt{JyNI.jar}, \texttt{libJyNI-Loader.so} and \texttt{libJyNI.so} on Jython's classpath.
\texttt{libJyNI-Loader.so} and \texttt{libJyNI.so} can alternatively be placed somewhere on the Java library path.
\end{itemize}

Jython should now be able to run \texttt{JyNIDemo.py} via%
\begin{quote}\begin{verbatim}
java -cp "[...]:JyNI.jar:[JyNI binaries folder]:
   jython.jar" org.python.util.jython JyNIDemo.py
\end{verbatim}

\end{quote}
Be sure to use Jython 2.7 (beta) or newer. If you are not using the Jython stand-alone version, make sure
that Jython's \texttt{Lib}-folder is on the Python path.

\subsection{Versioning note%
  \label{versioning-note}%
}

JyNI's version consists of two parts. The first part (currently 2.7) indicates the targeted API version. Your Jython
should meet this version if you intend to use it with JyNI. For extensions, the API version means that
a production release of JyNI would be able to load any native extension that a CPython distribution of the
same version (and platform) can load.
Of course, this is an idealistic goal – there will always remain some edgy, maybe exotic API-aspects JyNI won't be
able to support.

The second part of the JyNI version (currently alpha.2.1) indicates the development status. As long as it contains
“alpha” or “beta”, one can't expect that the targeted API version is already met. Once out of beta, we will maintain
this version part as a third index of the targeted API version (i.e. JyNI 2.7.x).

\section{Capabilities%
  \label{capabilities}%
}

JyNI is currently available for Linux only. Once it is sufficiently complete and stable, we will work out a cross platform version compilable on Windows, Mac OS X and others.
The following built-in types are already supported:%
\begin{itemize}

\item 

Number types \texttt{PyInt}, \texttt{PyLong}, \texttt{PyFloat}, \texttt{PyComplex}
\item 

Sequence types \texttt{PyTuple}, \texttt{PyList}, \texttt{PySlice}, \texttt{PyString}, \texttt{PyUnicode}
\item 

Data structure types \texttt{PyDict}, \texttt{PySet}, \texttt{PyFrozenSet}
\item 

Operational types \texttt{PyModule}, \texttt{PyClass}, \texttt{PyMethod}, \texttt{PyInstance}, \texttt{PyFunction}, \texttt{PyCode}, \texttt{PyCell}
\item 

Singleton types \texttt{PyBool}, \texttt{PyNone}, \texttt{PyEllipsis}, \texttt{PyNotImplemented}
\item 

Native types \texttt{PyCFunction}, \texttt{PyCapsule}, \texttt{PyCObject}
\item 

Natively defined custom types
\item 

Exception types
\item 

\texttt{PyType} as static type or heap type
\end{itemize}

The function families \texttt{PyArg\_ParseTuple} and \texttt{Py\_BuildValue} are also supported.

\section{Implementation%
  \label{implementation}%
}

To create JyNI we took the source code of CPython 2.7 and stripped away all functionality that can be provided by Jython and is not needed for mirroring objects (see below). We kept the interface unchanged and implemented it to delegate calls to Jython via JNI and vice versa.
The most difficult thing is to present JNI \texttt{jobject} s from Jython to extensions such that they look like \texttt{PyObject*} from Python (C-API). For this task, we use the three different approaches explained below, depending on the way a native type is implemented.

In this section, we assume that the reader is familiar with the Python \cite{C-API} and has some knowledge about the C programming language, especially about the meaning of pointers and memory allocation.

\subsection{Python wraps Java%
  \label{python-wraps-java}%
}

The best integration with Jython is obtained, if \texttt{PyObject*} is only a stub that
delegates all its calls to Jython (figure \DUrole{ref}{pwj}). This is only possible, if Jython features a
suitable counterpart of the \texttt{PyObject} (i.e. some subclass of \texttt{org.python.core.PyObject}
with similar name, methods and functionality).

Further, there must not exist macros
in the official C-API that directly access the \texttt{PyObject}'s memory. Consequently, one
cannot use \texttt{tp\_dictoffset} to obtain the object's dictionary or \texttt{offset} from
\texttt{PyMemberDef} to access the object's members.

Since members are usually only accessed via generic
getter or setter methods that also look for a \texttt{PyGetSetDef} with the right name, we usually re-implement
the members as get-sets. Also the dictionary access is usually performed in methods we can safely
rewrite to versions that get the dictionary from Jython.\begin{figure}[]\noindent\makebox[\columnwidth][c]{\includegraphics[scale=0.35]{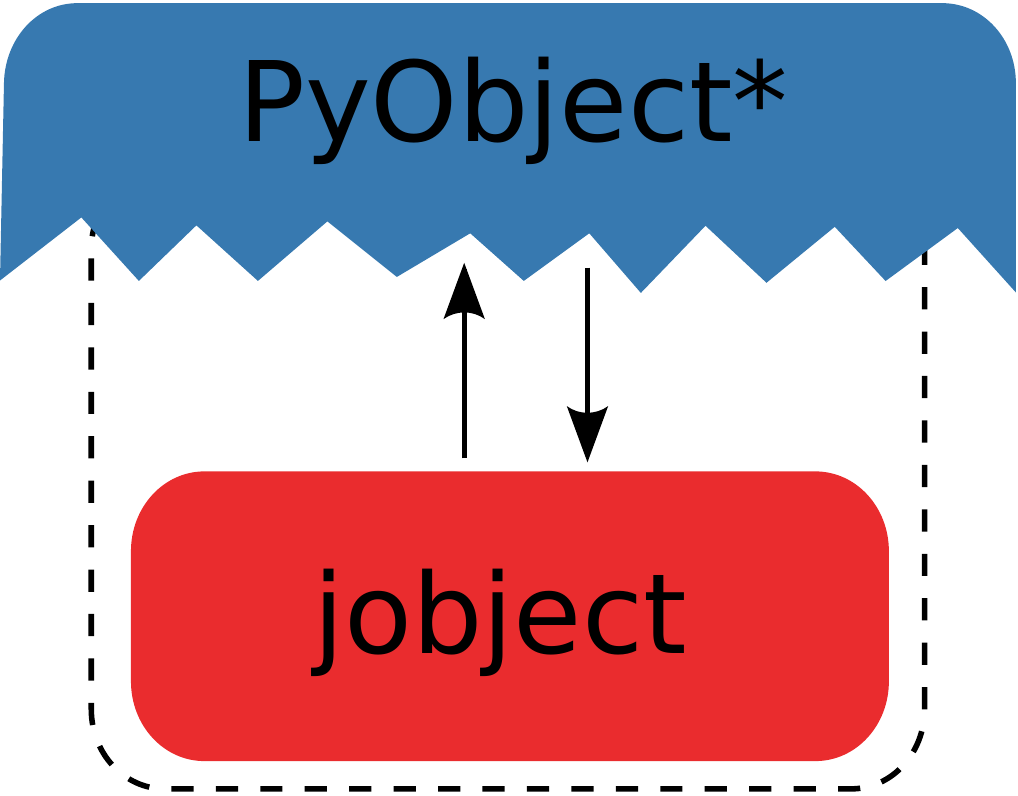}}
\caption{Python wraps Java \DUrole{label}{pwj}}
\end{figure}

Examples for this method are
\texttt{PyDict}, \texttt{PySlice} and \texttt{PyModule}.

The cases where this approach fails are%
\begin{itemize}

\item 

if Jython features no corresponding type
\item 

if the Python C-API features macros to access the Object's memory directly
\end{itemize}

We deal with these cases in the following.

\subsection{Mirroring objects%
  \label{mirroring-objects}%
}

If the Python C-API provides macros to access an object's data, we cannot setup
the object as a stub, because the stub would not provide the memory positions needed
by the macros. To overcome this issue, we mirror the object if its C-API features
such direct access macros (figure \DUrole{ref}{miro}).\begin{figure}[]\noindent\makebox[\columnwidth][c]{\includegraphics[scale=0.35]{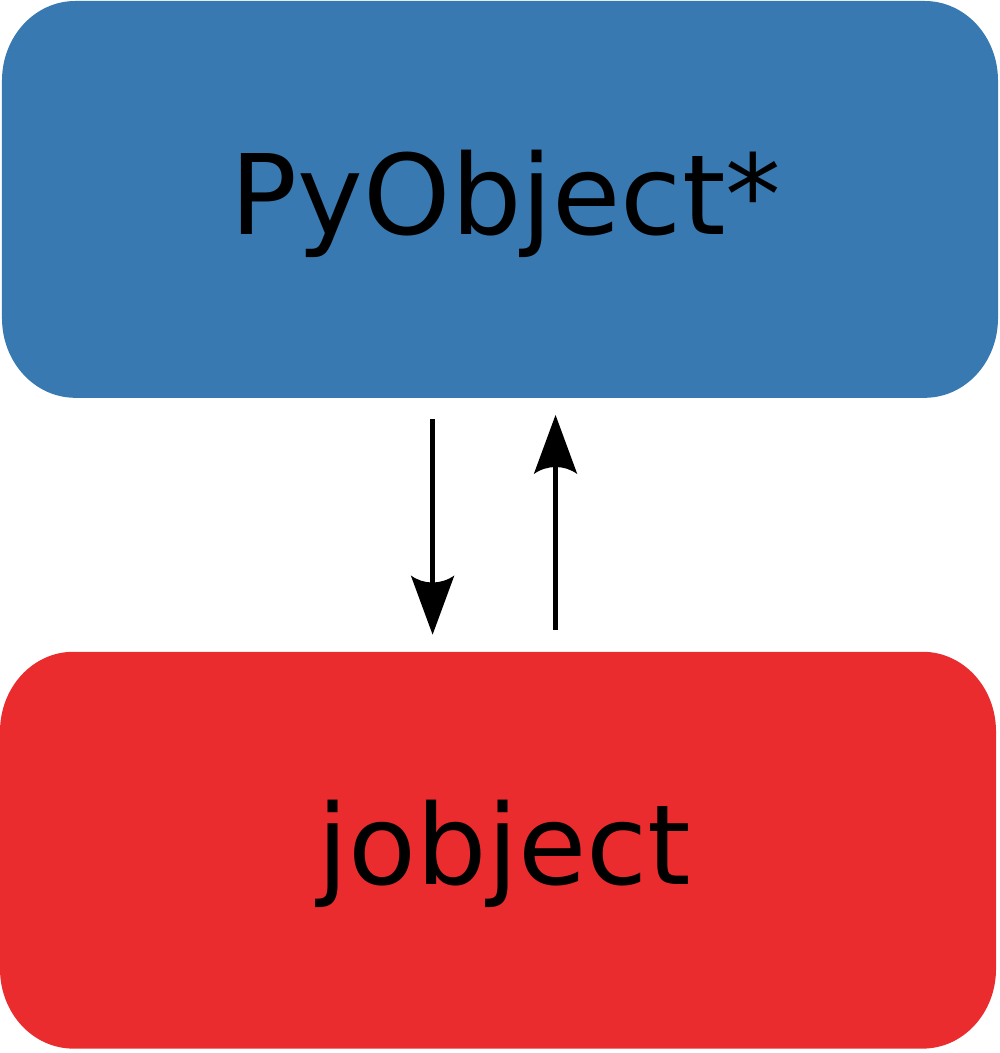}}
\caption{Objects are mirrored \DUrole{label}{miro}}
\end{figure}

Examples, where this approach is successfully applied are \texttt{PyTuple}, \texttt{PyList}, \texttt{PyString}, \texttt{PyInt}, \texttt{PyLong}, \texttt{PyFloat} and \texttt{PyComplex}.

The difficulty here is to provide a suitable synchronization between the counterparts.
If the CPython object is modified by C code, these changes must be reflected immediately on Jython side.
The problem here is, that such changes are not reported; they must be detected. Performing the synchronization when the C call returns to Jython is only suitable, if no multiple threads exist.
However, most of the affected objects are immutable anyway, so an initial data synchronization is sufficient.

\texttt{PyList} is an example for an affected object that is mutable via a macro. For \texttt{PyList}, we
perform an individual solution. The Jython class \texttt{org.python.core.PyList} uses a variable of type \texttt{java.util.List} (which is an interface) to store its backend. We wrote a wrapper, that provides access to the memory of the C struct of \texttt{PyListObject} and implements the \texttt{java.util.List} interface on Java side. If a \texttt{PyList} is mirrored, we replace its backend by our wrapper. If it was initially created on the Jython side, we insert all its elements into the C counterpart on initialization.

\texttt{PyCell} and \texttt{PyByteArray} are other examples that need mirror mode, but are mutable. However, we have rough ideas how to deal with them, but since they are not used by NumPy, we don't put priority on implementing them.

\subsection{Java wraps Python%
  \label{java-wraps-python}%
}

If Jython provides no counterpart of an object type, the two approaches described above are not feasible.
Typically, this occurs, if an extension natively defines its own \texttt{PyType} objects, but there are also examples for this in the original Python C-API. If the types were previously known, we could simply implement Jython counterparts for them and apply one of the two approaches above. However, we decided to avoid implementing new Jython objects if possible and solve this case with a general approach.
\texttt{PyCPeer} extends \texttt{org.python.core.PyObject} and redirects the basic methods to a native \texttt{PyObject*} (figure \DUrole{ref}{jwp}).
The corresponding \texttt{PyObject*} pointer is tracked as a java \texttt{long} in \texttt{PyCPeer}. Currently \texttt{PyCPeer} supports attribute access by delegating \texttt{\_\_findattr\_ex\_\_}, which is the backend method for all attribute accessing methods in Jython (i.e. \texttt{\_\_findattr\_\_} and \texttt{\_\_getattr\_\_} in all variants). Further, \texttt{PyCPeer} delegates the methods \texttt{\_\_str\_\_}, \texttt{\_\_repr\_\_} and \texttt{\_\_call\_\_}. A more exhaustive support is planned. \texttt{PyCPeerType} is a special version of \texttt{PyCPeer} that is suited to wrap a natively defined \texttt{PyType}.

Let's go through an example. If you execute the Python code \textquotedbl{}\texttt{x = foo.bar}\textquotedbl{},
Jython compiles it equivalently to the Java call \textquotedbl{}\texttt{x = foo.\_\_getattr\_\_(\textquotedbl{}bar\textquotedbl{});}\textquotedbl{}. If \texttt{foo} is a \texttt{PyCPeer} wrapping a native \texttt{PyObject*}, Java's late binding would call \texttt{\_\_findattr\_ex\_\_(\textquotedbl{}bar\textquotedbl{})} implemented in \texttt{PyCPeer}. Via the native method \texttt{JyNI.getAttrString(long peerHandle, String name)} the call is delegated to \texttt{JyNI\_getAttrString} in \texttt{JyNI.c} and then finally to \texttt{PyObject\_GetAttrString} in \texttt{object.c}. To convert arguments and return values between Java \texttt{jobject} and CPython \texttt{PyObject*}, we use the conversion methods \texttt{JyNI\_JythonPyObject\_FromPyObject} and \texttt{JyNI\_PyObject\_FromJythonPyObject} (see next section). Our version of \texttt{PyObject\_GetAttrString} falls back to the original CPython implementation, if it is called with a \texttt{PyCPeer} or a mirrored object. A flag in the corresponding \texttt{JyObject} (see next section) allows to detect these cases.\begin{figure}[]\noindent\makebox[\columnwidth][c]{\includegraphics[scale=0.35]{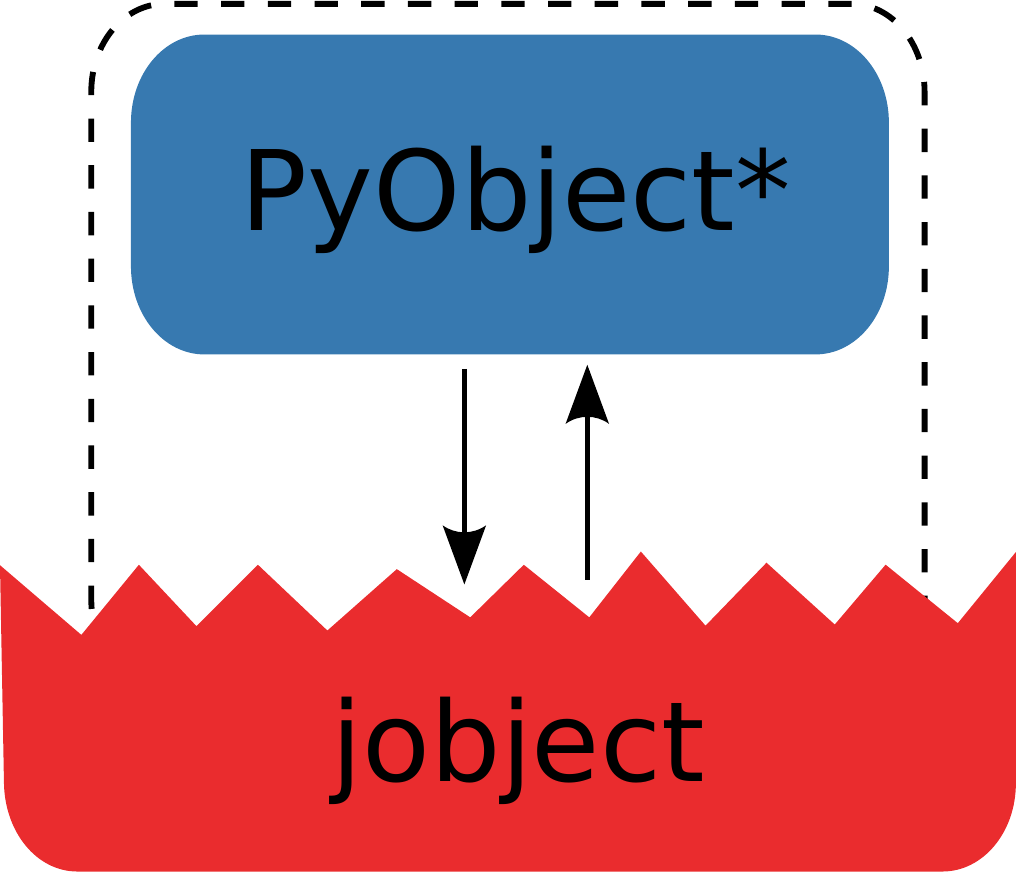}}
\caption{Java wraps Python \DUrole{label}{jwp}}
\end{figure}

An example from the C-API that needs the approach from this section is \texttt{PyCFunction}.

\subsection{Object lookup%
  \label{object-lookup}%
}

Every mentioned approach involves tying a \texttt{jobject} to a \texttt{PyObject*}. To resolve this connection
as efficiently as possible, we prepend an additional header before each \texttt{PyObject} in memory.
If a \texttt{PyGC\_Head} is present, we prepend our header even before that, as illustrated in figure \DUrole{ref}{objl}.\begin{figure}[]\noindent\makebox[\columnwidth][c]{\includegraphics[scale=0.35]{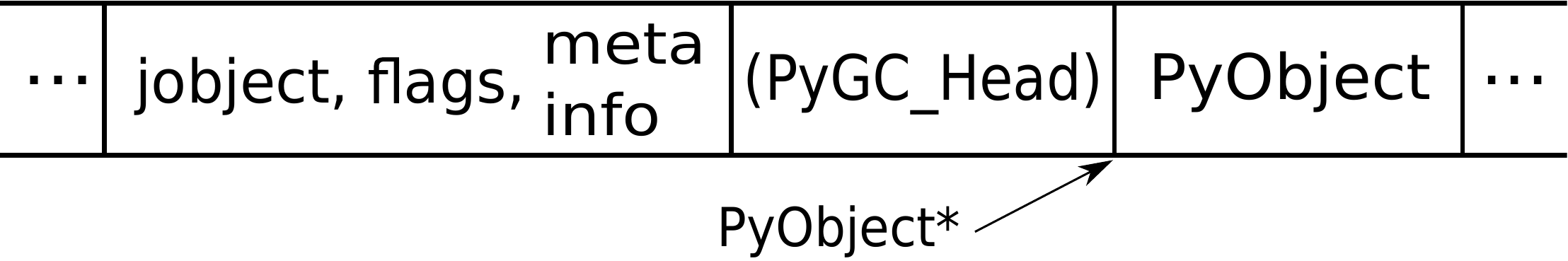}}
\caption{Memory layout \DUrole{label}{objl}}
\end{figure}

In the source, this additional header is called \texttt{JyObject} and defined as follows:\begin{Verbatim}[commandchars=\\\{\},fontsize=\footnotesize]
\PY{k}{typedef} \PY{k}{struct}
\PY{p}{\PYZob{}}
   \PY{n}{jobject} \PY{n}{jy}\PY{p}{;}
   \PY{k+kt}{unsigned} \PY{k+kt}{short} \PY{n}{flags}\PY{p}{;}
   \PY{n}{JyAttribute}\PY{o}{*} \PY{n}{attr}\PY{p}{;}
\PY{p}{\PYZcb{}} \PY{n}{JyObject}\PY{p}{;}
\end{Verbatim}
\texttt{jy} is the corresponding \texttt{jobject}, \texttt{flags} indicates which of the above mentioned approaches is used, whether a \texttt{PyGC\_Head} is present, initialization state and synchronization behavior.
\texttt{attr} is a linked list containing \texttt{void} pointers for various purpose. However, it
is intended for rare use, so a linked list is a sufficient data structure with minimal overhead. A \texttt{JyObject} can use it to save pointers to data that must be deallocated along with the \texttt{JyObject}. Such pointers typically arise when formats from Jython must be converted to a version that the original
\texttt{PyObject} would have contained natively.

To reserve the additional memory, allocation is adjusted wherever it occurs, e.g. when allocations inline as is the case for number types. The adjustment also occurs in \texttt{PyObject\_Malloc}. Though this method might not only be used for \texttt{PyObject} allocation, we always prepend space for a \texttt{JyObject}. We regard this slight overhead in non-\texttt{PyObject} cases as preferable over potential segmentation fault if a \texttt{PyObject} is created via \texttt{PyObject\_NEW} or \texttt{PyObject\_NEW\_VAR}.
For these adjustments to apply, an extension must be compiled with the \texttt{WITH\_PYMALLOC} flag activated.
Otherwise several macros would direct to the raw C methods \texttt{malloc}, \texttt{free}, etc., where the neccessary
extra memory would not be reserved. So an active \texttt{WITH\_PYMALLOC} flag is crucial for JyNI to work.
However, it should be not much effort to recompile affected extensions with an appropriate \texttt{WITH\_PYMALLOC} flag value.

Statically defined \texttt{PyType} objects are treated as a special case, as their memory is not dynamically allocated. We resolve them simply via a lookup table when converting from \texttt{jobject} to \texttt{PyObject*} and via a name lookup by Java reflection if converting the other way. \texttt{PyType} objects dynamically allocated on the heap are of course not subject of this special case and are treated like usual \texttt{PyObject} s (the \texttt{Py\_TPFLAGS\_HEAPTYPE} flag indicates this case).

The macros \texttt{AS\_JY(o)} and \texttt{FROM\_JY(o)}, defined in \texttt{JyNI.h}, perform the necessary pointer arithmetics to get the \texttt{JyObject} header from a \texttt{PyObject*} and vice versa. They are not intended for direct use, but are used internally by the high-level conversion functions described below, as these also consider special cases like singletons or \texttt{PyType} objects.

The other lookup direction is done via a hash map on the Java side. JyNI stores the \texttt{PyObject*} pointers as Java \texttt{Long} objects and looks them up before doing native calls. It then directly passes the pointer to the native method.

The high-level conversion functions\begin{Verbatim}[commandchars=\\\{\},fontsize=\footnotesize]
\PY{n}{jobject} \PY{n+nf}{JyNI\PYZus{}JythonPyObject\PYZus{}FromPyObject}
   \PY{p}{(}\PY{n}{PyObject}\PY{o}{*} \PY{n}{op}\PY{p}{)}\PY{p}{;}
\PY{n}{PyObject}\PY{o}{*} \PY{n+nf}{JyNI\PYZus{}PyObject\PYZus{}FromJythonPyObject}
   \PY{p}{(}\PY{n}{jobject} \PY{n}{jythonPyObject}\PY{p}{)}\PY{p}{;}
\end{Verbatim}
take care of all this, do a lookup and automatically perform initialization if the lookup fails.
Of course the \texttt{jobject} mentioned in these declarations must not be an arbitrary \texttt{jobject}, but one that extends \texttt{org.python.core.PyObject}.
Singleton cases are also tested and processed appropriately. \texttt{NULL} converts to \texttt{NULL}.
Though we currently see no use case for it, one can use the declarations in \texttt{JyNI.h} as JyNI C-API. With the conversion methods one could write hybrid extensions that do C, JNI and Python calls natively.

\subsection{Global interpreter lock (GIL)%
  \label{global-interpreter-lock-gil}%
}

The global interpreter lock is a construction in CPython that prevents multiple threads from running Python code in the same process. It is usually acquired when the execution of a Python script begins and released when it ends. However, a native extension and some parts of native CPython code can release and re-acquire it by inserting the \texttt{Py\_BEGIN\_ALLOW\_THREADS} and \texttt{Py\_END\_ALLOW\_THREADS} macros. This way, an extension can deal with multiple threads and related things like input events (f.i. Tkinter needs this).

In contrast to that, Jython does not have a GIL and allows multiple threads at any time, using Java's threading architecture. Since native extensions were usually developed for CPython, some of them might rely on the existence of a GIL and might produce strange behaviour if it was missing. So JyNI features a GIL to provide most familiar behaviour to loaded extensions. To keep the Java parts of Jython GIL-free and have no regression to existing multithreading features, the JyNI GIL is only acquired when a thread enters native code and released when it enters Java code again – either by returning from the native call or by performing a Java call to Jython code. Strictly speaking, it is not really global (thus calling it “GIL” is a bit misleading), since it only affects threads in native code. While there can always be multiple threads in Java, there can only be one thread in native code at the same time (unless the above mentioned macros are used).

\section{A real-world example: Tkinter%
  \label{a-real-world-example-tkinter}%
}

To present a non-trivial example, we refere to Tkinter, one of the most popular GUI frameworks for Python.
There has already been an approach to make Tkinter available in Jython, namely jTkinter – see \cite{JTK}. However,
the last update to the project was in 2000, so it is rather outdated by now and must be considered inactive.

Since release alpha.2.1, JyNI has been tested successfully on basic Tkinter code.
We load Tkinter from the place where it is usually installed on Linux:\begin{Verbatim}[commandchars=\\\{\},fontsize=\footnotesize]
\PY{k+kn}{import} \PY{n+nn}{sys}
\PY{c}{\PYZsh{}Include native Tkinter:}
\PY{n}{sys}\PY{o}{.}\PY{n}{path}\PY{o}{.}\PY{n}{append}\PY{p}{(}\PY{l+s}{\PYZsq{}}\PY{l+s}{/usr/lib/python2.7/lib\PYZhy{}dynload}\PY{l+s}{\PYZsq{}}\PY{p}{)}
\PY{n}{sys}\PY{o}{.}\PY{n}{path}\PY{o}{.}\PY{n}{append}\PY{p}{(}\PY{l+s}{\PYZsq{}}\PY{l+s}{/usr/lib/python2.7/lib\PYZhy{}tk}\PY{l+s}{\PYZsq{}}\PY{p}{)}

\PY{k+kn}{from} \PY{n+nn}{Tkinter} \PY{k+kn}{import} \PY{o}{*}

\PY{n}{root} \PY{o}{=} \PY{n}{Tk}\PY{p}{(}\PY{p}{)}
\PY{n}{txt} \PY{o}{=} \PY{n}{StringVar}\PY{p}{(}\PY{p}{)}
\PY{n}{txt}\PY{o}{.}\PY{n}{set}\PY{p}{(}\PY{l+s}{\PYZdq{}}\PY{l+s}{Hello World!}\PY{l+s}{\PYZdq{}}\PY{p}{)}

\PY{k}{def} \PY{n+nf}{print\PYZus{}text}\PY{p}{(}\PY{p}{)}\PY{p}{:}
    \PY{k}{print} \PY{n}{txt}\PY{o}{.}\PY{n}{get}\PY{p}{(}\PY{p}{)}

\PY{k}{def} \PY{n+nf}{print\PYZus{}time\PYZus{}stamp}\PY{p}{(}\PY{p}{)}\PY{p}{:}
    \PY{k+kn}{from} \PY{n+nn}{java.lang} \PY{k+kn}{import} \PY{n}{System}
    \PY{k}{print} \PY{l+s}{\PYZdq{}}\PY{l+s}{System.currentTimeMillis: }\PY{l+s}{\PYZdq{}}
        \PY{o}{+}\PY{n+nb}{str}\PY{p}{(}\PY{n}{System}\PY{o}{.}\PY{n}{currentTimeMillis}\PY{p}{(}\PY{p}{)}\PY{p}{)}

\PY{n}{Label}\PY{p}{(}\PY{n}{root}\PY{p}{,}
 \PY{n}{text}\PY{o}{=}\PY{l+s}{\PYZdq{}}\PY{l+s}{Welcome to JyNI Tkinter\PYZhy{}Demo!}\PY{l+s}{\PYZdq{}}\PY{p}{)}\PY{o}{.}\PY{n}{pack}\PY{p}{(}\PY{p}{)}
\PY{n}{Entry}\PY{p}{(}\PY{n}{root}\PY{p}{,} \PY{n}{textvariable}\PY{o}{=}\PY{n}{txt}\PY{p}{)}\PY{o}{.}\PY{n}{pack}\PY{p}{(}\PY{p}{)}
\PY{n}{Button}\PY{p}{(}\PY{n}{root}\PY{p}{,} \PY{n}{text}\PY{o}{=}\PY{l+s}{\PYZdq{}}\PY{l+s}{print text}\PY{l+s}{\PYZdq{}}\PY{p}{,}
            \PY{n}{command}\PY{o}{=}\PY{n}{print\PYZus{}text}\PY{p}{)}\PY{o}{.}\PY{n}{pack}\PY{p}{(}\PY{p}{)}
\PY{n}{Button}\PY{p}{(}\PY{n}{root}\PY{p}{,} \PY{n}{text}\PY{o}{=}\PY{l+s}{\PYZdq{}}\PY{l+s}{print timestamp}\PY{l+s}{\PYZdq{}}\PY{p}{,}
            \PY{n}{command}\PY{o}{=}\PY{n}{print\PYZus{}time\PYZus{}stamp}\PY{p}{)}\PY{o}{.}\PY{n}{pack}\PY{p}{(}\PY{p}{)}
\PY{n}{Button}\PY{p}{(}\PY{n}{root}\PY{p}{,} \PY{n}{text}\PY{o}{=}\PY{l+s}{\PYZdq{}}\PY{l+s}{Quit}\PY{l+s}{\PYZdq{}}\PY{p}{,}
            \PY{n}{command}\PY{o}{=}\PY{n}{root}\PY{o}{.}\PY{n}{destroy}\PY{p}{)}\PY{o}{.}\PY{n}{pack}\PY{p}{(}\PY{p}{)}

\PY{n}{root}\PY{o}{.}\PY{n}{mainloop}\PY{p}{(}\PY{p}{)}
\end{Verbatim}
\begin{figure}[]\noindent\makebox[\columnwidth][c]{\includegraphics[scale=0.40]{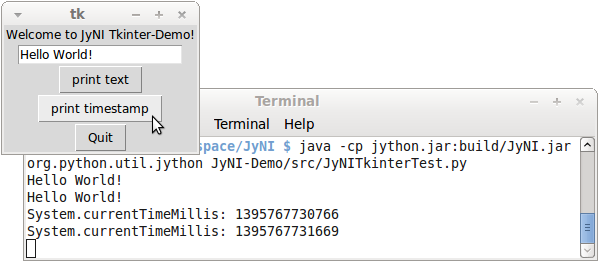}}
\caption{Tkinter demonstration \DUrole{label}{tkDemo}}
\end{figure}Note that the demonstration also runs with CPython in principle. To make this
possible, we perform \texttt{from java.lang import System} inside the method body
of \texttt{print\_time\_stamp} rather than at the beginning of the file. Thus, only the
button “print timestamp” would produce an error, since it performs Java calls.
In a Jython/JyNI environment, the button prints the current output of
\texttt{java.lang.System.currentTimeMillis()} to the console (see figure \DUrole{ref}{tkDemo}).

\section{Another example: The \texttt{datetime} module%
  \label{another-example-the-datetime-module}%
}

As a second example, we refere to the \texttt{datetime} module. Jython features a Java-based version of that module, so this does not yet pay off in new functionality.
However, supporting the original native \texttt{datetime} module is a step toward NumPy,
because it features a public C-API that is needed by NumPy. The following code demonstrates how JyNI can load the original \texttt{datetime} module. Note that we load it
from the place where it is usually installed on Linux. To overwrite the Jython version,
we put the new path to the beginning of the list in \texttt{sys.path}:\begin{Verbatim}[commandchars=\\\{\},fontsize=\footnotesize]
\PY{k+kn}{import} \PY{n+nn}{sys}
\PY{n}{sys}\PY{o}{.}\PY{n}{path}\PY{o}{.}\PY{n}{insert}\PY{p}{(}\PY{l+m+mi}{0}\PY{p}{,} \PY{l+s}{\PYZsq{}}\PY{l+s}{/usr/lib/python2.7/lib\PYZhy{}dynload}\PY{l+s}{\PYZsq{}}\PY{p}{)}
\PY{k+kn}{import} \PY{n+nn}{datetime}
\PY{k}{print} \PY{n}{datetime}\PY{o}{.}\PY{n}{\PYZus{}\PYZus{}doc\PYZus{}\PYZus{}}
\PY{k}{print} \PY{l+s}{\PYZdq{}}\PY{l+s}{\PYZhy{}}\PY{l+s}{\PYZdq{}} \PY{o}{*} \PY{l+m+mi}{22}
\PY{k}{print}

\PY{k}{print} \PY{n}{datetime}\PY{o}{.}\PY{n}{\PYZus{}\PYZus{}name\PYZus{}\PYZus{}}
\PY{n}{now} \PY{o}{=} \PY{n}{datetime}\PY{o}{.}\PY{n}{datetime}\PY{p}{(}\PY{l+m+mi}{2013}\PY{p}{,} \PY{l+m+mi}{11}\PY{p}{,} \PY{l+m+mi}{3}\PY{p}{,} \PY{l+m+mi}{20}\PY{p}{,} \PY{l+m+mi}{30}\PY{p}{,} \PY{l+m+mi}{45}\PY{p}{)}

\PY{k}{print} \PY{n}{now}
\PY{k}{print} \PY{n+nb}{repr}\PY{p}{(}\PY{n}{now}\PY{p}{)}
\PY{k}{print} \PY{n+nb}{type}\PY{p}{(}\PY{n}{now}\PY{p}{)}
\end{Verbatim}
To verify that the original module is loaded, we print out the \texttt{\_\_doc\_\_} string. It must read \textquotedbl{}Fast implementation of the datetime type.\textquotedbl{}. If JyNI works as excpected, the
output is:%
\begin{quote}\begin{verbatim}
Fast implementation of the datetime type.
----------------------

datetime
2013-11-03 20:30:45
datetime.datetime(2013, 11, 3, 20, 30, 45)
<type 'datetime.datetime'>
\end{verbatim}

\end{quote}

\section{Roadmap%
  \label{roadmap}%
}
The main goal of JyNI is compatibility with NumPy and SciPy, since these extensions are of most scientific importance.
Since NumPy has dependencies on several other extensions, we will have to ensure compatibility with these extensions first.
Among these are ctypes and datetime (see previous section). In order to support ctypes, we will have to support the \texttt{PyWeakRef} object.

\subsection{Garbage Collection%
  \label{garbage-collection}%
}

Our first idea to provide garbage collection for native extensions, was to adopt the original CPython garbage collector source and use it in parallel with the Java garbage collector.
The CPython garbage collector would be responsible to collect mirrored objects, native stubs and objects created by native extensions. The stubs would keep the corresponding objects alive by maintaining a global reference. However, this approach does not offer a clean way to trace reference cycles through Java/Jython code (even pure Java Jython objects can be part of reference cycles keeping native objects alive forever).

To obtain a cleaner solution, we plan to setup an architecture that makes the native objects subject to Java's garbage collector. The difficulty here is that Java's mark and sweep algorithm only traces Java objects. When a Jython object is collected, we can use its finalizer to clean up the corresponding C-\texttt{PyObject} (mirrored or stub), if any. To deal with native \texttt{PyObject} s that don't have a corresponding Java object, we utilize \texttt{JyGCHead} s (some minimalistic Java objects) to track them and clean them up on finalization. We use the visitproc mechanism of original CPython's garbage collection to obtain the reference connectivity graph of all relevant native \texttt{PyObject} s. We mirror this connectivity in the corresponding \texttt{JyGCHead} s, so that the Java garbage collector marks and sweeps them according to native connectivity.

A lot of care must be taken in the implementation details of this idea. For instance, it is not obvious, when to update the connectivity graph. So a design goal of the implementation is to make sure that an outdated connectivity graph can never lead to the deletion of still referenced objects. Instead, it would only delay the deletion of unreachable objects. Another issue is that the use of Java finalizers is discouraged for various reasons. An alternative to finalizers are the classes from the package \texttt{java.lang.ref}. We would have \texttt{JyGCHead} extend \texttt{PhantomReference} and register all of them to a \texttt{ReferenceQueue}. A deamon thread would be used to poll references from the queue as soon as the garbage collector enqueues them. For more details on Java reference classes see \cite{JREF}.

\subsection{Cross-Platform support%
  \label{cross-platform-support}%
}

We will address cross-platform support when JyNI has reached a sufficiently stable state on our development platform.
At least we require rough solutions for the remaining gaps. Ideally, we focus
on cross-platform support when JyNI is capable of running NumPy.

\section{License%
  \label{license}%
}

JyNI is released under the GNU \cite{GPL} version 3.
To allow for commercial use, we add the classpath exception \cite{GPL_EXC} like known from GNU Classpath to it.

%
%

We were frequently asked, why not LGPL, respectively what the difference to LGPL is.
In fact, the GPL with classpath exception is less restrictive than LGPL.
\cite{GPL_EXC} states this as follows:
The LGPL would additionally require you to \textquotedbl{}allow modification of the portions of the library you use\textquotedbl{}.
For C/C++ libraries this especially requires distribution of the compiled .o-files from the pre-linking stage.
Further you would have to allow \textquotedbl{}reverse engineering (of your program and the library) for debugging such modifications\textquotedbl{}.

\end{document}

%% file: page_numbers.tex
\setcounter{page}{59}

%% file: paper.bbl
\begin{thebibliography}{IRONCLAD}
\bibitem[JyNI]{JyNI}{

Stefan Richthofer, Jython Native Interface (JyNI) Homepage, \url{http://www.JyNI.org}, 6 Apr. 2014, Web. 7 Apr. 2014}
\bibitem[JYTHON]{JYTHON}{

Python Software Foundation, Corporation for National Research Initiatives, Jython: Python for the Java Platform, \url{http://www.jython.org}, Mar. 2014, Web. 7 Apr. 2014}
\bibitem[IRONCLAD]{IRONCLAD}{

Resolver Systems, Ironclad, \url{http://code.google.com/p/ironclad}, 26 Aug. 2010, Web. 7 Apr. 2014}
\bibitem[CPYEXT]{CPYEXT}{

PyPy team, PyPy/Python compatibility, \url{http://pypy.org/compat.html}, Web. 7 Apr. 2014}
\bibitem[NP4J]{NP4J}{

Joseph Cottam, NumPy4J, \url{https://github.com/JosephCottam/Numpy4J}, 11. Dec. 2013, Web. 7 Apr. 2014}
\bibitem[JEPP]{JEPP}{

Mike Johnson, Java embedded Python (JEPP), \url{http://jepp.sourceforge.net}, 14 May 2013, Web. 7 Apr. 2014}
\bibitem[JTK]{JTK}{

Finn Bock, jTkinter, \url{http://jtkinter.sourceforge.net}, 30 Jan. 2000, Web. 7 Apr. 2014}
\bibitem[C-API]{C-API}{

Python Software Foundation, Python/C API Reference Manual, \url{http://docs.python.org/2/c-api}, Web. 7 Apr. 2014}
\bibitem[JREF]{JREF}{

Peter Haggar, IBM Corporation, \url{http://www.ibm.com/developerworks/library/j-refs}, 1 Oct. 2002, Web. 7 Apr. 2014}
\bibitem[GPL]{GPL}{

Free Software Foundation, GNU General Public License v3, \url{http://www.gnu.org/licenses/gpl.html}, 29 June 2007, Web. 7 Apr. 2014}
\bibitem[GPL\_EXC]{GPL_EXC}{

Wikipedia, GPL linking exception, \url{http://en.wikipedia.org/wiki/GPL_linking_exception\#The_classpath_exception}, 5 Mar 2014, Web. 7 Apr. 2014}
\end{thebibliography}
